# Direct imaging of topological edge states at a bilayer graphene domain wall

Long-Jing Yin[1], Hua Jiang[2], Jia-Bin Qiao[1], Lin He[1,*]

[1]Center for Advanced Quantum Studies, Department of Physics, Beijing Normal University, Beijing, 100875, People's Republic of China

[2]College of Physics, Optoelectronics and Energy, Soochow University, Suzhou 215006, People's Republic of China

*Correspondence to: helin@bnu.edu.cn.

**Abstract**: The AB-BA domain wall in gapped graphene bilayers is a rare naked structure hosting topological electronic states. Here we show, for the first time, direct imaging of its topological edge states by using scanning tunneling microscope. The simultaneously obtained atomic-resolution images of the domain wall provide us unprecedented opportunities to measure the spatially-varying edge states within it. The one-dimensional conducting channels are observed to be mainly located around the two edges of the domain wall, which is reproduced quite well by our theoretical calculations. Our experiment further demonstrates that the one-dimensional topological states are quite robust even in the presence of high magnetic fields.

**One Sentence Summary:** The one-dimensional symmetry-protected topological states are directly observed, for the first time, in the AB-BA domain wall of gapped bilayer graphene by using scanning tunneling microscope.

**Main Text:** Looking for systems where topological edge states persist in the absence of external magnetic fields boosts rapid developments in condensed matter physics in the past few years (*1-14*). Gapped graphene bilayer with smooth domain walls is predicted to be one of the most promising candidates where charge carriers can travel long distances without dissipation (*8-12*). The domain wall separating two oppositely biased bilayer graphene is first proposed by Martin et al. to host one-dimensional (1D) topological states (*8*). Later, the domain wall between AB- and BA-stacked bilayer graphene under a uniform external field is demonstrated to be equivalent to the gate-polarity domain wall (*8*) and it is believed to be a crystalline topological defect hosting symmetry-protected topological gapless mode because of a change in the Chern number (*12*). Very recently, the existence of topologically protected 1D chiral states have been demonstrated explicitly in the two types of domain walls through transport measurement (*13,14*), opening up opportunities for exploring unique topological states in graphene bilayer.

The AB-BA domain wall in graphene bilayer, with electrons residing right at the surface, provides unprecedented opportunities to directly image the topologically protected 1D conducting channels (Fig. 1A and Fig. 1B). More importantly, such a crystalline topological line defect exists naturally in Bernal graphene bilayers grown by chemical vapor deposition (*15,16*) and in exfoliated bilayer graphene (that is, prepared using adhesive type) from graphite (*13*). Here, we report, for the first time, direct imaging of the topologically protected 1D conducting channels in the AB-BA domain wall in exfoliated graphene bilayer. The exfoliated bilayer and trilayer graphene flakes were deposited on the substrate (here the supporting substrate is graphite) during the process of mechanical exfoliation and, very importantly, these graphene sheets decouple from the graphite surface due to the presence of the stacking misorientation with the underlying substrates, as demonstrated in this paper and in previous studies (*17-24*).

To identify the AB-BA domain wall in decoupled bilayer graphene on graphite, we used both the STM images and the scanning tunneling spectroscopy (STS) spectra. Firstly, the decoupled bilayer graphene on graphite exhibits a small period of moiré patterns (that is, with a large rotation angle with the substrate) in the STM measurements (*18,20,24*) and its atomic-resolution STM image shows a triangular lattice because of the A/B atoms' asymmetry in the topmost Bernal bilayer (see Fig. S1). The decoupled monolayer graphene also exhibits a small period of moiré patterns, however, its atomic-resolution STM images show a hexagonal lattice (see Fig. S1). The high-field STS spectra provide further information about the stacking orders of the topmost few layers (*21*): the decoupled Bernal bilayer shows Landau quantization of massive Dirac Fermions (Fig. 2 and Fig. S2) (*21,25*), whereas, the decoupled monolayer exhibits Landau quantization of massless Dirac fermions (see Fig. S1) (*17,18*).

Once the decoupled bilayer graphene is identified, we used STM measurements to find 1D structures (see Fig. 1C as an example) in the bilayer region as a possible candidate for the AB-BA domain wall. The strong dependence of the 1D structure on the bias voltage (used for imaging), as shown in Fig. S3, excludes the graphene nanoripple (*26*) or nanowrinkle (*27*) as the origin of the 1D structure. We attribute the 1D structure in Fig. 1C to the AB-BA domain wall in bilayer graphene. The reason that we can observe the AB-BA domain walls in the STM measurement owing to their relatively higher conductivity comparing to that of the adjacent gapped bilayer regions (see Fig. 2F). The ultra-low random potential fluctuations due to substrate imperfections allows us to obtain high-quality atomic-resolution STM images of the domain wall, as shown in Fig. 1C and Fig. 1D. We obtained a triangular lattice, as is characteristic of Bernal-stacked bilayer, both in the left and right regions, whereas, we obtained a hexagonal-like lattice in the center of the domain wall (see the insets of Fig. 1C). The 1D structure with hexagonal-like lattice at its center and AB and BA domains surrounding directly demonstrated that the studied structure is the AB-BA domain wall. Fig. 1D shows a representative atomic-resolution STM image of the AB-BA domain wall. From left to right of the domain wall, the two graphene sheets translate relative to each other in opposite directions (one translates downward, the other translates upward), completing an interlayer translation from AB to BA stacking. The interatomic distances in the domain wall and in the Bernal bilayer regions were further analysed by taking a 2D Fourier transform and the interatomic distances in the domain wall are $(1.5 \pm 0.5)\%$ smaller than those in the surrounding Bernal regions (Fig. S4). To complete a one-bond-length armchair-direction interlayer translation from AB to BA stacking, the width of the domain wall is estimated to be about 9.0 nm, which agrees quite well with the measured value $\sim (8.0 \pm 1.0)$ nm. The angle between the boundary normal and the translation direction is measured to be about 85°, indicating that the studied domain wall is almost a purely shear soliton (*15*).

The electronic properties around the AB-BA domain wall are further studied by STS measurements, as shown in Fig. 2, Fig. S5, and Fig. S6. The spectra recorded in both the AB- and BA-stacked regions exhibit characteristics that are expected to be observed in gapped graphene bilayers (*21,25,28*). The substrate breaks the inverse symmetry of the topmost adjacent bilayers and then a finite gap $\sim 80$ meV is generated in the parabolic bands of the Bernal bilayer (the charge neutrality points of the two Bernal bilayer regions are measured to differ about 15 meV). At the level of low-energy effective theory, the AB-stacked bilayer is equivalent to the BA-stacked bilayer subjecting to the opposite gate polarity (*12*). Thus, the sign of the energy gap changes across the domain wall from the AB- to BA-stacked regions, and symmetry-protected gapless modes are expected to emerge in the domain wall. In high magnetic fields, the spectra recorded in the Bernal bilayer regions exhibit Landau quantization of massive Dirac fermions

(Fig. 2D, Fig. 2E and see Supplementary for further analysis). The two lowest Landau levels (LLs) $LL_{(0,1,+)}$ and $LL_{(0,1,-)}$ (here 0/1 are Landau indices and +/- are valley indices), which are a couple of layer-polarized quartets, depends on the sign of the gate polarity (or the sign of the energy gap) of the two Bernal bilayer regions (*28*). Therefore, they are reversed in the adjacent AB- and BA-stacked regions, as shown in Fig. 2A and Fig. 2B. In the experiment, the measured local density of states (LDOS) at position *r* are determined by the wavefunctions, while the wavefunctions of LLs have their spatial extent, $\sim 2\sqrt{N}l_B$ (here *N* is the Landau index and $l_B = \sqrt{\hbar/eB}$, which is of the order of 10 nm for the magnetic fields applied in our experiment) (*25*). Consequently, we can detect Landau levels of both the AB and BA domains in the spectra recorded in the AB-BA domain wall (Fig. 2C). The "splitting" of the Landau levels recorded in the domain wall, as shown in Figs. 2C and 2G, arises from the relatively shift of the charge neutrality points of the adjacent AB and BA domains. By using similar STM measurements, layer stacking domain walls in trilayer graphene, which separate ABA- and ABC-stacked trilayer graphene, have also been observed unambiguously in our experiment (see Fig. S7 and Fig. S8 for an example). Our result demonstrates that the layer stacking domain walls naturally exist in graphene multilayers and affect their electronic properties dramatically.

To further confirm the existence of symmetry-protected topological conducting channels in the AB-BA domain wall, we directly imaged these 1D states by operating energy-fixed STS mapping, which reflects the LDOS in real space. Fig. 3A-C shows several STS maps at different energies. At the energies within the band gap of the adjacent AB and BA domains, clearly 1D conducting channels can be observed along the domain wall. A notable feature of the topological states is that they mainly located at the two edges of the AB-BA domain wall and such a feature is independent of the energy of the gapless edge states. To verify the spatial distribution of the gapless states, we calculated electronic structures of a shear domain wall with a finite width *W* (see Fig. 3D for an example). Fig. 3E shows a schematic representation of the domain wall. In the calculation, we consider a tight-binding Hamiltonian with nearest-neighbor intra- and interlayer hopping and a finite chemical potential difference between two layers is taking into account to describe the energy gap observed in the Bernal regions (*12*). The symmetry-protected gapless edge states emerge in the domain wall (Fig. 3D), which is irrespective of the type and width of the domain wall (see Supplementary for details of calculation). In our experiment, the STS maps probe predominantly the LDOS of the top layer. To compare with the experimental result, we plot a theoretical spatial-distribution of the gapless edge states in the topmost graphene layer in Fig. 3E. Obviously, the topological states are mainly located at the two edges of the domain wall and this feature is found to be independent of the probed energy within the gap of the Bernal bilayer regions. Here we should point out that such a spatial distribution of the topological states is independent of the edges of the domain wall (see Fig. S9). Therefore, our experimental observations are reproduced quite well by the theoretical calculations. This provides direct and compelling evidence that the symmetry-protected topological edge states exist in the AB-BA domain walls of gapped bilayer graphene.

The STS maps of the gapless edge states are also measured in the presence of high magnetic fields (Fig. 3F). It is remarkable that these states are quite robust even in the highest magnetic field ~ 8 T of our STM system. More importantly, the FWHM (full-width at half-maximum) of topological states along the two edges of the domain wall decreases with increasing the magnetic fields, which may further diminish any possible scattering of the topological edge states along the AB-BA domain walls. In a very recent transport measurement, it was also demonstrated that the topological feature of the gapless edge states is very robust against the perturbation of external magnetic fields and the backscattering of the topological states is further suppressed in the presence of magnetic fields (*14*). Our work thus demonstrates the robust feature of the symmetry-protected topological edge states in the AB-BA domain walls of gapped bilayer graphene, opening a wide vista of graphene-based topological transport properties.

**Acknowledgments:** This work was supported by the National Basic Research Program of China (Grants Nos. 2014CB920903, 2013CBA01603, 2014CB920901), the National Natural Science Foundation of China (Grant Nos. 11422430, 11374035, 11374219), the program for New Century Excellent Talents in University of the Ministry of Education of China (Grant No. NCET-13-0054), Beijing Higher Education Young Elite Teacher Project (Grant No. YETP0238). L.H. also acknowledges support from the National Program for Support of Top-notch Young Professionals.

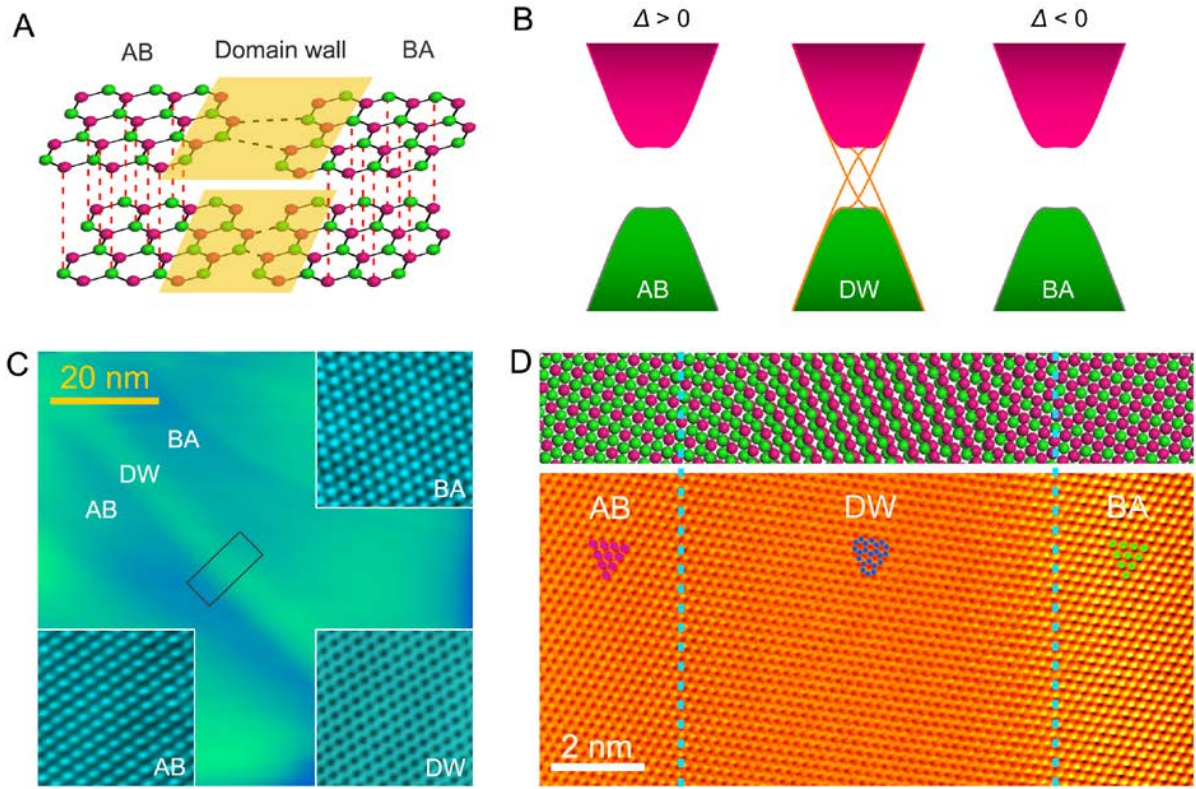

**Fig. 1. AB-BA domain wall in bilayer graphene.** (**A**) Schematic representation of an AB-BA domain wall in bilayer graphene. (**B**) Schematic band structures of the AB, domain wall (DW) and BA regions of a bilayer graphene. The AB- and BA-stacked regions are gapped. The topological edge states (orange curves) emerge in the DW region. (**C**) 80 nm × 80 nm STM topographic image of a decoupled bilayer graphene region on graphite surface ($V_b$ = 0.4 V, $I$ = 0.25 nA). An AB-BA domain wall is observed in the bilayer. Insets: atomic-resolution STM images in the AB, DW and BA regions, respectively. (**D**) A typical atomic-resolution STM current image (lower panel) across the AB-BA domain wall (rectangular region in **C**). A schematic image of the domain wall is shown in the upper panel. A transition from triangular lattice (in the AB region) to hexangular-like lattice (in the center of the domain wall) and then to triangular lattice (in the BA region) is clearly observed. The width of the domain wall is estimated to be (8 ± 1) nm.

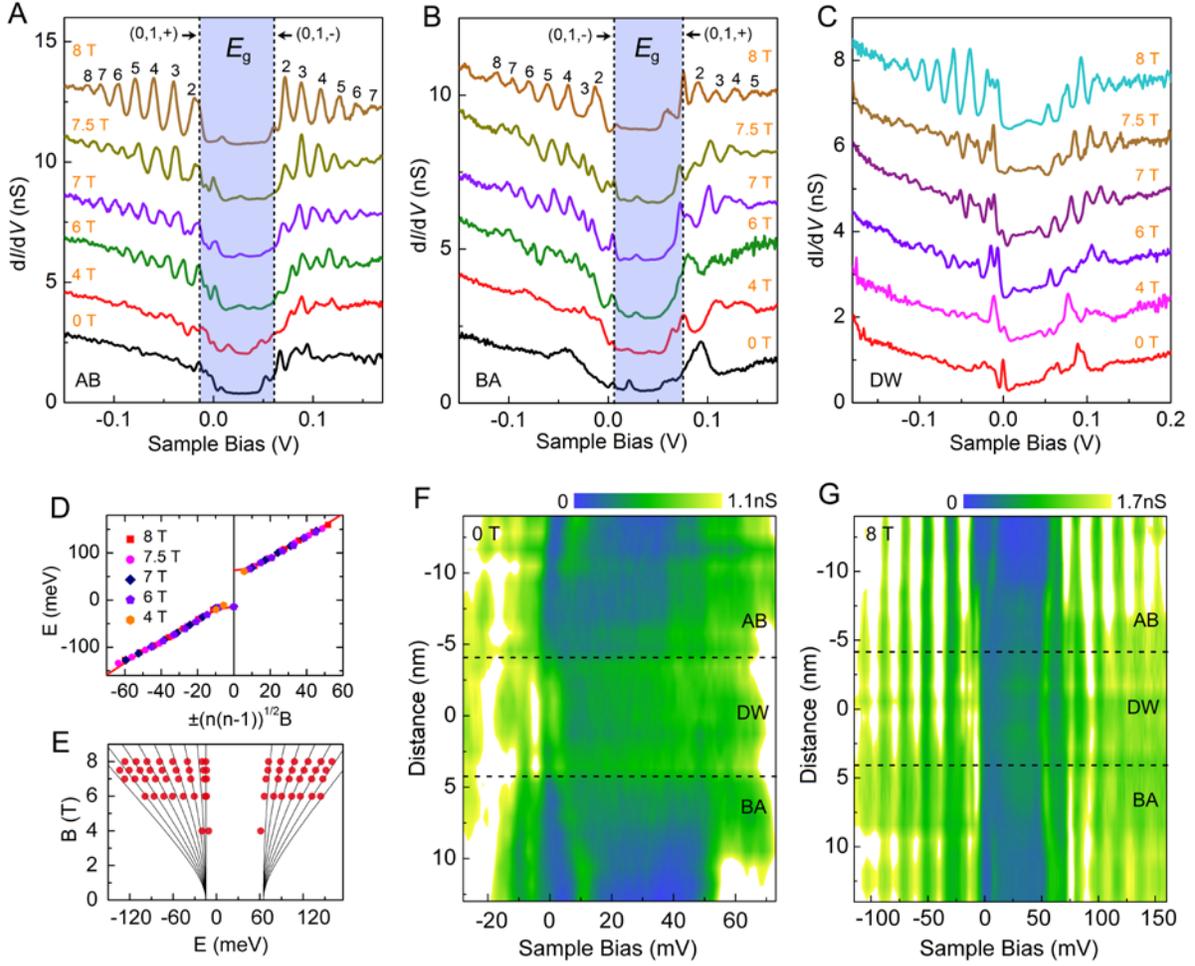

**Fig. 2. Microscopic properties of the AB-BA domain wall.** (**A** to **C**) Tunneling spectra of the gaped graphene bilayers recorded at the AB-stacked region (**A**), the BA-stacked region (**B**) and the domain wall region (**C**) under various magnetic fields. LL peak indices are marked (± are valley indices) and the gap are labeled by shadows in the AB and BA bilayer regions. The tunneling curves are offset in y-axis for clarity. (**D** and **E**) LL peaks energies extracted from (A) plotted versus $\pm(n(n-1))^{1/2}B$ (D) and the magnetic fields $B$ (E). The solid curves are the fitting of the data with the theoretical equation [Eq. (S1)] yielding the band gap of $E_g = 80 \pm 1$ meV and effective mass of $m^* = (0.0454 \pm 0.0001)m_e$ ($m_e$ is the free-electron mass). (**F** and **G**) STS spectra maps at 0 T (**F**) and 8 T (**G**) measured across the AB-BA domain wall. The zero-position is defined at the middle of the domain wall.

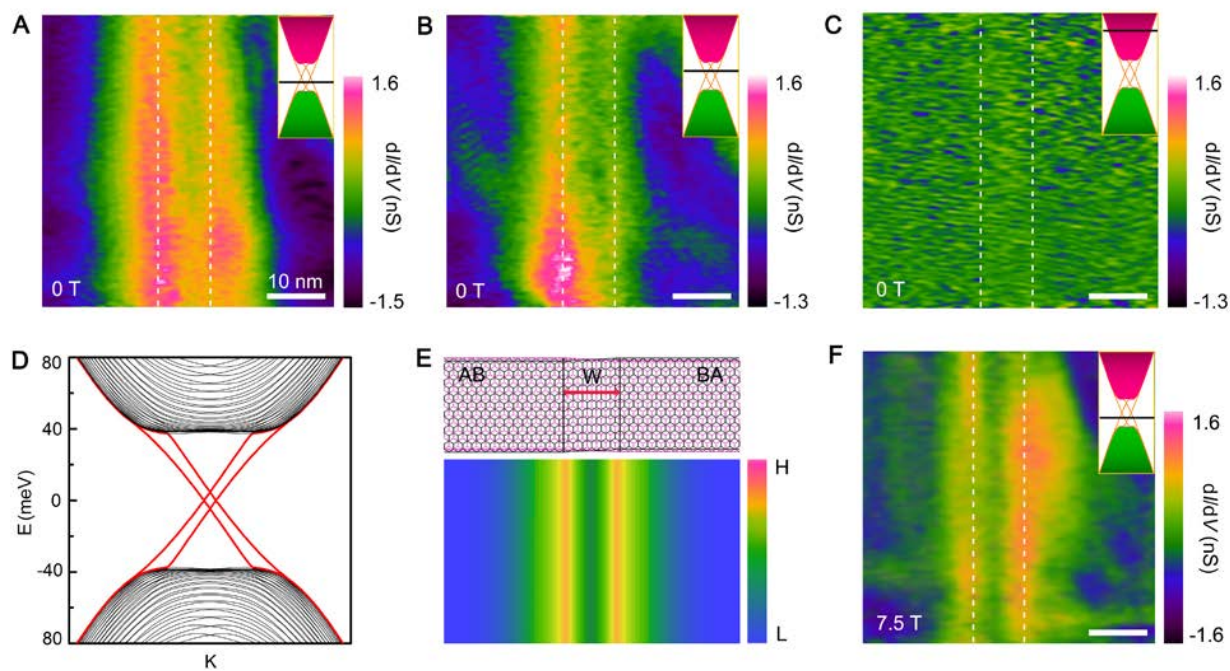

**Fig. 3. Direct imaging of the 1D conducting channels at the AB-BA domain wall.** (**A** to **C**) d*I*/d*V* maps recorded under 0 T along the AB-BA domain wall with the fixed sample bias of 30 mV (**A**), 40 mV (**B**) and 300 mV (**C**), respectively. The 1D topological states are predominantly located at the two edges of the domain wall. (**D**) A representative theoretical band structure of an AB-BA domain wall with width of 8 nm and the gap in the Bernal region of 80 meV. (**E**) Upper: illustration of an AB-BA domain wall. Lower: spatial distribution of the topological states around the domain wall obtained by theoretical calculation. (**F**) STS map of the domain wall taken at 7.5 T with sample bias of 30 mV. The two edges of the AB-BA domain wall are labeled by dashed lines.

**Supplementary Materials:**

Materials and Methods

Landau quantization in gapped bilayer graphene

Calculation of the topological edge states in the AB-BA domain walls

Figures S1-S9
References S1-S3